\documentclass[twocolumn]{aastex631}

\usepackage{amsmath}
\usepackage{color}
\usepackage{float}
\usepackage{bm}
\usepackage{soul}
\usepackage{xcolor}
\usepackage{orcidlink}

\shorttitle{Polarization ratios}
\shortauthors{Polanco-Rodríguez, Krafft \& Savoini}

\begin{document}

\title{Polarization ratios of turbulent Langmuir/$\mathcal{Z}$-mode waves generated by electron beams \\ in magnetized solar wind plasmas}

\correspondingauthor{Catherine Krafft}
\email{catherine.krafft@universite-paris-saclay.fr}

\author{F.J. Polanco-Rodríguez\orcidlink{0009-0005-2951-697X}}
\affiliation{Laboratoire de Physique des Plasmas (LPP), CNRS, Sorbonne Université, \\ 
Observatoire de Paris, Université Paris-Saclay,\\ Ecole polytechnique, Institut Polytechnique de Paris, 91120 Palaiseau, France}

\author{C. Krafft\orcidlink{0000-0002-8595-4772}}
\affiliation{Laboratoire de Physique des Plasmas (LPP), CNRS, Sorbonne Université, \\ 
Observatoire de Paris, Université Paris-Saclay,\\ Ecole polytechnique, Institut Polytechnique de Paris, 91120 Palaiseau, France}

\affiliation{Institut Universitaire de France (IUF)}

\author{P. Savoini\orcidlink{0000-0002-2117-3803}}
\affiliation{Laboratoire de Physique des Plasmas (LPP), CNRS, Sorbonne Université, \\ 
Observatoire de Paris, Université Paris-Saclay,\\ Ecole polytechnique, Institut Polytechnique de Paris, 91120 Palaiseau, France}

\begin{abstract}
 
The polarization ratios $F=|E_\perp|^2/|E|^2$ of beam-generated turbulent Langmuir/$\mathcal{Z}$-mode ($\mathcal{LZ}$)  waves and electromagnetic emissions radiated at plasma frequency  $\omega_p$ by such sources are studied in weakly magnetized and randomly inhomogeneous plasmas owing to large-scale and long-term 2D/3V Particle-In-Cell simulations  with  parameters relevant to type III solar radio bursts.  Statistical studies using waveforms recorded by virtual satellites are performed to determine the distributions of polarization ratios as a function of  beam and plasma parameters. This efficient method, which mimics waveform recording by spacecraft in the solar wind, leads to results consistent with observations. Moreover, plasma random density fluctuations $\delta n$ turn out to be the key factor responsible for the increase in polarization ratios  up to $F\simeq 1$. Indeed, it is demonstrated that linear mode conversion at constant frequency of $\mathcal{LZ}$ waves scattering on  $\delta n$ is the most  efficient and fast process to produce large polarization ratios in randomly inhomogeneous plasmas, due to electromagnetic slow extraordinary $\mathcal Z$-mode wave emission by $\mathcal{LZ}$ wave turbulence.  Results  provide guidance to theoretical studies and useful support to estimate the average level of density fluctuations $\Delta N$ in solar wind plasmas.  
\end{abstract}

\section{Introduction}
Since decades, a large amount of high- and low-frequency waveforms are observed at different solar wind conditions by various spacecraft (e.g. \textcite{HospodarskyGurnett1995}, \textcite{Kellogg1999}, \textcite{MalaspinaErgun2008}, \textcite{Malaspina2011}, \textcite{Krasnoselskikh2011}, \cite{GrahamCairns2013,GrahamCairns2014}, \textcite{Soucek2021}, \textcite{Pisa2021}, \textcite{Graham2021}, \textcite{Raouafi2023}, \textcite{Lorfing2023}, and references therein) and, more recently, by the Solar Orbiter and Parker Solar Probe missions (\cite{Fox2016}, \cite{Muller2020}). 
Their analysis has shed light on dispersion, polarization and excitation mechanisms of various types of waves.

In particular, electric field waveforms of Langmuir/$\mathcal{Z}$-mode waves (hereinafter referred to as $\mathcal{LZ}$ waves) excited by electron beams propagating along open magnetic field lines during type III solar radio bursts are commonly observed.
Most of them reveal predominant parallel (to the ambient magnetic field $\mathbf{B}_0$) electric field component $E_{\parallel}$.
However, a significant proportion of waves with large perpendicular electric fields $E_{\perp}$ and high polarization ratios $F=\sum_t E_{\perp}(t)^{2}/\sum_t (E_{\perp}(t)^{2}+E_{\parallel}(t)^{2})$ are also reported (\textcite{Bale1996}, \textcite{Malaspina2011}, \textcite{Kellogg2013}, \textcite{GrahamCairns2013,GrahamCairns2014}). 
Their origin, which remains unclear, is linked to the major problem of electromagnetic radiation by electrostatic wave turbulence generated by beams. 
Therefore, the questions of which wave processes and beam-plasma parameters are responsible for such large polarization ratios are addressed in this work. 

Several hypotheses have been proposed to explain why polarization ratios calculated using measured $\mathcal{LZ}$ waveforms can reach large values up to $F\sim0.4$ and even more (\textcite{Malaspina2011}, \textcite{Graham2012}, \textcite{GrahamCairns2013,GrahamCairns2014}). 
Electrostatic wave decay was invoked, as it transports the energy of $\mathcal{LZ}$ waves towards smaller $k$-scales while enhancing their electromagnetic character (\textcite{WillesCairns2000}, \textcite{GrahamCairns2013}, \textcite{Layden2013}, \textcite{Polanco2025}), as well as three-dimensional Langmuir eigen modes (\textcite{MalaspinaErgun2008}, \textcite{Ergun2008}) or linear transformations of $\mathcal{LZ}$ waves on density gradients (\textcite{Krauss-Varban1989}, \textcite{Bale1998}, \textcite{Kellogg1999}, \textcite{Malaspina2011}).

Random density fluctuations $\delta n$ exist in the solar wind up to average levels $\Delta N=\langle (\delta n/n_0)^2\rangle^{1/2}\simeq 0.07$, where $n_{0}$ is the ambient plasma density (e.g. \textcite{Celnikier1987}, \textcite{Krupar2020}). 
Recently, the magnetic signature of an electromagnetic slow extraordinary $\mathcal{Z}$-mode wave was observed for the first time in a solar wind plasma with strong density fluctuations (\cite{Larosa2022}), and is thought to be produced via linear mode conversion (LMC)  at constant frequency of $\mathcal{LZ}$ waves scattering on $\delta n$ (e.g. \cite{Hinkel-Lipsker1989}, \cite{Krasnoselskikh2019}, \cite{KrafftSavoini2022a}). 
Recently, this process has been studied analytically and numerically in  randomly inhomogeneous and weakly magnetized two- and three-dimensional plasmas (\cite{Krafft2025}), demonstrating the predominance of $\mathcal{Z}$-mode wave radiation by beam-generated radio sources. 
In this view, it is now timely to explain the link between $\mathcal{Z}$-mode waves and large polarization ratios. 

The main questions raised in this work are the following. 
What is the main process responsible for large wave polarization ratios $F$ in solar wind plasmas? How are they distributed  for a statistical ensemble of  $\mathcal{LZ}$ and electromagnetic wavepackets?  What is the link between large polarization ratios and the generation of electromagnetic $\mathcal{Z}$-mode waves? What is the impact of beam-plasma parameters (magnetization, temperatures, average level of density fluctuations, beam velocity) on $F$ distributions? How can  properties of polarization ratios be used  to diagnose solar wind plasmas? 

These issues are addressed by using large-scale and long-term 2D/3V Particle-In-Cell (PIC) simulations with  parameters relevant to type III solar radio bursts.
In this context, we also demonstrate the effectiveness of the technique used  to analyze the simulations. 
It consists in the generation of a large number of virtual satellites measuring waveforms (\cite{Krafft2014}) during their motion through a large 2D simulation box. 
To our knowledge, such approach is used for the first time in the framework of PIC simulations, and is accompanied here by statistical studies using several thousands of waveforms. 
This method, which mimics wave measurements by space-born satellites, aims to support the analysis and interpretation of  solar wind observations. 
Indeed, long-term and high resolution waveforms can be obtained simultaneously for the six fields' components as well as for the ion and electron densities, considerably increasing  the possibility of identifying ongoing processes and wave characteristics, compared to the situation in space.
Moreover, it is shown that this local approach is particularly useful for randomly inhomogeneous plasmas where waves are strongly scattered and for which the global approach previously used (e.g. \cite{KrafftSavoini2024}) provides, for some diagnostics, less clear and interpretable results.

\section{Methods}
\subsection{2D/3V PIC simulations}\label{section PIC Simulation}

Large-scale and long-term 2D/3V PIC simulations are performed using the SMILEI code (\textcite{Derouillat2018}), with beam and plasma parameters relevant to type III solar radio bursts  at distances of $\sim0.1-1$ au from the Sun.
Randomly inhomogeneous and weakly to moderately magnetized plasmas are considered, with $\Delta N\leq0.05$ and ratios of cyclotron to plasma frequency in the range $\omega_c/\omega_p\leq 0.5$.
An electron beam with drift velocity $v_b=12.7v_T\simeq0.25c$  ($v_T$ is the electron thermal velocity) and weak relative density $n_b=5\cdot10^{-4}n_0$ is injected in the 2D plasma along the ambient magnetic field $\mathbf{B}_0$ directed along the $x$-axis.
The electron-to-ion mass and temperature ratios are $m_e/m_i=1/1836$ and $T_e/T_i=10$.

Two other sets of parameters are used to understand the dependence of polarization ratios  on $\Delta N$,  $\omega_c/\omega_p$,  $T_e$, $T_e/T_i$ and $v_b$.
The first one,  with $v_b=12.7v_T\simeq0.25c$ like mentioned above, is characterized by a  low electron temperature $T_e=20eV$, with $T_e/T_i=3$, $\omega_c/\omega_p=0.02$ and  $\Delta N=$0,   $0.025$. 
The second one includes slower beams with $6v_T \leq v_b\leq 12.7v_T$ (i.e. $0.12c \leq v_b\leq 0.25c$) and plasma parameters $0.025\leq\Delta N\leq0.05$, $T_e/T_i=10$ and $\omega_c/\omega_p=0.07$.

Simulation planes with sizes $L_x\times L_y=1448\times1448\lambda_D^2$ are used to perform exceptionally long-term simulations up to $t=30,000\omega_p^{-1}$, which can describe the full dynamics of electrostatic decay in the presence of a weak beam. 
Note that the large number  of particles per cell used ($N_c=5400$) reduces the total energy relative error to $\sim10^{-4}$ at $t=30,000\omega_p^{-1}$. 
More details can be found in previous articles by the authors (\cite{KrafftSavoini2021,KrafftSavoini2022a,KrafftSavoini2022b,KrafftSavoini2023,KrafftSavoini2024}, \cite{Krafft2024,Krafft2025}, \cite{Polanco2025}).

\subsection{Virtual satellites}
Hereafter we simulate thousands of virtual satellites moving in  a 2D simulation box  and recording at each time $t$ and position $\mathbf{x}(t)$ the electric and magnetic wave fields $\mathbf{E} (\mathbf{x}(t),t)$ and $\mathbf{B}(\mathbf{x}(t),t)$, as well as the ion and electron densities $n_{i}(\mathbf{x}(t),t)$ and $n_{e}(\mathbf{x}(t),t)$, with a sampling rate ranging from 0.06$\omega_p$ to 0.18$\omega_p$.
For purpose of simplicity and without loss of generality, the velocity $\mathbf{v}_{s}$ of each satellite with respect to the plasma considered as immobile is chosen constant and is directed along the $x$-axis, i.e. along $\mathbf{v}_{b}$ and $\mathbf{B}_{0}$, in order to simplify the determination of the Doppler shift. 
Its modulus $v_{s}=0.3v_{T}$ is a nominal value for spacecraft moving relative to the slow solar wind (e.g. \cite{GrahamCairns2013}). However, it can be adjusted to facilitate waveforms' analysis, e.g.  to separate  from each other, in Doppler-shifted power spectra, the different groups of $\mathcal{LZ}$ waves structured by electrostatic decay (beam-driven, forward- and back-scattered) in weakly inhomogeneous plasmas with $\Delta N\gtrsim 0$. 

The duration of the simulated waveforms is $T=15,000-30,000\omega_p^{-1}$, which is roughly equivalent to $0.5-1$ seconds in the solar wind near 1 au. 
Wave frequencies measured by the virtual satellites are Doppler-shifted, i.e. $\omega^D=\omega-\mathbf k\cdot\mathbf v_s$, where $\omega$ and $\mathbf{k}$ are the frequency and the wavevector in the plasma (laboratory) frame, respectively. 
For ordinary and extraordinary electromagnetic waves radiated at $\omega_p$ (i.e. $\mathcal{O}$, $\mathcal{X}$ and $\mathcal{Z}$-mode waves), the Doppler shift term $\mathbf k\cdot\mathbf v_s$ is negligible (for the satellite velocity chosen) compared to their frequencies $\omega_{em}$ in the motionless plasma frame, as their wavenumbers satisfy $k\lambda_D\lesssim0.01$, so that $\omega^D_{em}\simeq\omega_{em}$. 
Frequencies $\omega_{\mathcal{S}}$ of ion acoustic waves in the plasma frame can be neglected compared to those in the satellite frame ($\omega^D_{\mathcal{S}}$) and to the Doppler shift term, so that $\omega^D_{\mathcal{S}}\simeq\pm \mathbf{k}_{\mathcal{S}} \cdot \mathbf{v_s}$. 
Finally, frequencies of $\mathcal{LZ}$ waves are Doppler shifted below or above $\omega_p$ depending on their forward or backward direction of propagation, respectively. 
For example, for beam-driven $\mathcal{LZ}$ waves with $ k_{\mathcal{LZ}}\lambda_D\sim 0.1$, we get $\omega^D_{\mathcal{LZ}}\simeq\omega_p(1+3k_{\mathcal{LZ}}^2\lambda_D^2/2+{\omega_c^2\sin^2\theta}/{2\omega_p^2})-v_{s}k_{x\mathcal{LZ}}\simeq\omega_p(1+3k_{\mathcal{LZ}}^2\lambda_D^2/2)-v_{s}k_{\mathcal{LZ}}\simeq 0.984$ for $\omega_c/\omega_p=0.07\ll 1$, where $k_{x\mathcal{LZ}}=\mathbf{k}_{\mathcal{LZ}}\cdot\mathbf{B}_0/B_0$.

\subsection{Tools for waveform analysis }

We present hereafter the methodology used to analyze the numerous waveforms recorded by virtual satellites, based on averaging methods and calculation of probability density functions (PDFs). The polarization ratio of a single waveform $w$ at time $t$ and frequency $\omega\simeq\omega_p$ is designated by 
\begin{equation}
\begin{split}
    F_{w}(t)=\frac{|E_{\perp}^w(\omega_p,t)|^2}{|E^w(\omega_p,t)|^2},  
\end{split}
\label{Gwt}
\end{equation}
where  $|E_{\perp}^w(\omega_p,t)|^2$ and $| E^w(\omega_p,t)|^2$ are the spectral energies  at $\omega\simeq\omega_p$, calculated using complex Morlet wavelet transforms, of the perpendicular and total electric fields of the waveform $w$, with $|E_{\perp}^w(\omega_p,t)|^2=|E_{\perp1}^w(\omega_p,t)|^2+|E_{\perp2}^w(\omega_p,t)|^2$  and  ($E_{\parallel}^w,E_{\perp 1}^w, E_{\perp 2}^w)=(E_x^w,E_y^w,E_z^w)$. 
We define the average of $F_w(t)$ over $N_w$ waveforms as
\begin{equation}
    \langle F(t)\rangle_w=\frac{1}{N_w}\sum_w F_w(t), \label{F2}
\end{equation}
whereas the average of $F_w(t)$ over the time interval $t_0\leq t\leq t_0+\Delta T$ is
\begin{equation}
    \langle F_w(t)\rangle_t=\frac{1}{\Delta T}\int_{t_0}^{t_0+\Delta T} F_w(t)dt. \label{F3}
\end{equation}
The polarization ratio averaged over time and waveforms is designated by $\langle F\rangle_{w,t}$. 
The PDF of $F_w(t)$ for a single waveform is $f_w(F)$, with $0\leq F\leq1$  representing $F_w(t)$ at a given time, whereas the PDF of $F_w(t)$ for $N_w $ waveforms is defined as $\langle f\rangle_w$. 
The distribution $\langle f(F,X_0)\rangle_w$ represents the PDF of  $N_w $ waveforms with polarization ratio $F$ at a given physical variable $X=X_0$.
Finally, since $F_w(t)$ is a ratio of energies, one can define a similar quantity in the $(\mathbf k,\omega)-$space as
\begin{equation}
\begin{split}
        F_{\mathbf k,\omega} =\frac{|E_{\perp \mathbf k,\omega}|^2}{|E_{ \mathbf k,\omega}|^2}, 
\end{split}
\label{Fk}
\end{equation}
where $E_{\perp  \mathbf k,\omega}$ and $E_{\mathbf k,\omega}$ are the Fourier components of $E_{\perp}(\mathbf x,t)$ and $ E(\mathbf x,t)$, respectively. 
$F_{\mathbf k,\omega}$ represents the polarization ratio of a wave with wavevector $\mathbf k$ and frequency $\omega$. 
Some correspondence between $\langle F\rangle_{w,t}$ and $F_{\mathbf k,\omega}$ can be established, since $\langle F\rangle_{w,t}$ is averaged over time and  space (each waveform depends on $(x(t),y)$).

\section{Waveforms and related diagnostics}
Before to perform statistical studies, let us analyze typical waveforms obtained by the simulations in a homogeneous and randomly inhomogeneous plasma.
\subsection{Homogeneous magnetized plasma}\label{section homogeneous magnetized plasma}
\begin{figure*}
    \centering
    \includegraphics[width=1\textwidth]{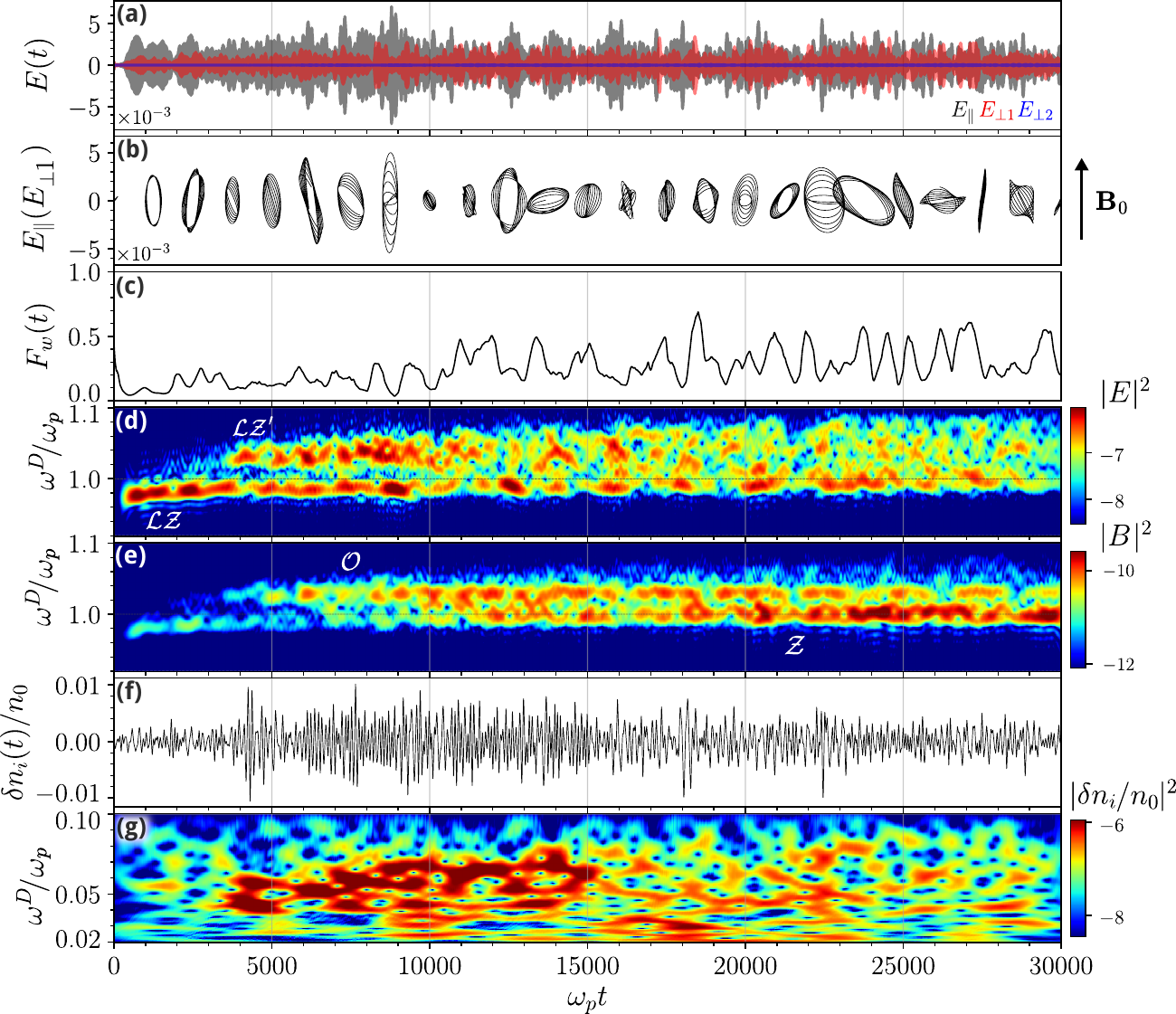}
    \caption{Typical example of waveform, together with hodograms, polarization ratios and wave energies, in a weakly magnetized homogeneous plasma with $\omega_c/\omega_p=0.07$ and $\Delta N=0$, for $0\leq\omega_p t\leq30,000$. 
    (a) Waveform of electric fields $E_\parallel(t)=E_x(t)$ (gray), $E_{\perp1}(t)=E_y(t)$ (red), and $E_{\perp2}(t)=E_z(t)$ (blue), filtered in the frequency interval $[0.9,1.1]\omega_p$. 
    (b) Hodograms $E_\parallel(E_{\perp1})$ calculated in equidistant time windows of $60\omega_p^{-1}$. 
    (c) Time variation of the polarization ratio $F_w(t)$ of the waveform. 
     (d) Electric energy spectrum $|E(\omega^D,t)|^2$ in the map ($\omega_pt,\omega^D/\omega_p$); the beam-driven $\mathcal{LZ}$  and backscattered $\mathcal{LZ^\prime}$ waves are indicated by labels.
    (e) Magnetic energy spectrum $|B(\omega^D,t)|^2$ in the map ($\omega_pt,\omega^D/\omega_p$);  $\mathcal{O}$- and $\mathcal{Z}$-modes are indicated by labels. Both spectra are calculated using wavelet transforms.
    (f) Variation of the ion density perturbation $\delta n_i(t)/n_0=(n_i(t)-n_0)/n_0$  as a function of  the time $\omega_pt$. 
    (g) Ion density perturbation energy spectrum $|\delta n_i(\omega^D,t)|^2$  in the map ($\omega_pt,\omega^D/\omega_p$), calculated using wavelet transform. 
    All fields and  energies  are presented in arbitrary units. 
    Color bars are in logarithmic scales.}
    \label{fig1}
\end{figure*}

Figure \ref{fig1} shows a typical waveform in a homogeneous magnetized plasma with $\omega_c/\omega_p=0.07$. 
It presents the electric fields $E_{\parallel}(t)$, $E_{\perp1}(t)$, and $E_{\perp2}(t)$, together with the hodograms $E_\parallel(E_{\perp1})$ calculated within time periods of $60\omega_p^{-1}$, the polarization ratio $F_w(t)$, the ion density perturbation $\delta n_i(t)/n_0=(n_i(t)-n_0)/n_0$, as well as the wavelet transforms $|E(\omega^D,t)|^2$,  $|B(\omega^D,t)|^2$ and $|\delta n_i(\omega^D,t)|^2$ of electric, magnetic and ion acoustic energies, respectively. 
Note that in Figures \ref{fig1}a-c, electric fields are filtered in the frequency range $[0.9,1.1]\omega_p$. 
The field $E_\parallel(t)$ remains the largest one  during most of time until $E_{\perp1}(t)$ reaches comparable amplitudes at large  $\omega_pt\gtrsim15,000$ (Figure \ref{fig1}a), with a constantly negligible contribution of $E_{\perp2}(t)$. 
In Figure \ref{fig1}d, the spectral electric energy $|E(\omega^D,t)|^2$ reaches large intensities at early times around $\omega^D\simeq0.98\omega_p$, corresponding to beam-driven $\mathcal{LZ}$ waves;
later, at $\omega_pt\gtrsim4000$, backward propagating waves $\mathcal{LZ}'$ produced by the first cascade $\mathcal{LZ}\longrightarrow\mathcal{LZ}'+\mathcal{S}'$ of electrostatic decay  (e.g. \cite{KrafftSavoini2024}) appear in the frequency range $1.01\lesssim\omega^D/\omega_p\lesssim 1.04$, simultaneously with ion acoustic wavepackets of frequencies $0.04\lesssim\omega^D/\omega_p\lesssim0.06$ (Figures \ref{fig1}f,g). 
$\mathcal{LZ}$ wave beatings correlated with ion acoustic oscillations are signatures of this process (e.g. in the range $3,500\lesssim\omega_pt\lesssim 5000$, see Figures \ref{fig1}a,d,f-g). 
Frequencies of forward and backward propagating  waves coming from the successive decay cascades, i.e.  $\mathcal{LZ}^{(2)}$ and  $\mathcal{LZ}^{(4)}$ (with $\omega^D\lesssim\omega_p$) and $\mathcal{LZ}^{\prime}$ and $\mathcal{LZ}^{(3)}$ (with $\omega^D\gtrsim\omega_p$), approach $\omega_p$ as time increases whereas, at earlier times $\omega_pt\lesssim10,000$, no energy is visible at $\omega_p$ (Figure \ref{fig1}d);  most of electric energy is concentrated around this frequency  at $\omega_pt\gtrsim20,000$, as a result of the ultimate stage of $\mathcal{LZ}$ wave decay. 

In Figure \ref{fig1}b, hodograms  are mainly linearly polarized and elongated along the magnetic field direction, as expected for quasi-parallel propagating electrostatic $\mathcal{LZ}$  waves; however, they present more complex structures as time increases, due to the occurrence of nonlinear wave processes.  
The polarization ratio $F_w(t)$ oscillates with time around the mean $\langle F_w(t)\rangle_t\simeq0.26$ (averaged over $0\lesssim\omega_pt\lesssim 30,000$), while slowly growing (Figure \ref{fig1}c). 
Enhancements up to $F_w(t)\simeq 0.6$ are observed at $\omega_p t\gtrsim 10,000$, in conjunction with the radiation of $\mathcal{Z}$-mode wave magnetic energy (Figure  \ref{fig1}e), resulting from the late stage of electrostatic decay when $\mathcal{LZ}$ wave energy has been transported down to electromagnetic $k$-scales. 
Indeed, the time variation of $|B(\omega^D,t)|^2$ exhibits a gradual increase of $\mathcal{Z}$-mode energy with frequencies $\omega^D\lesssim\omega_p$. 
At $\omega_p t\gtrsim 10,000$, magnetic signatures of $\mathcal{LZ}$ waves with  $\omega^D\gtrsim\omega_p$ become visible in Figure \ref{fig1}e; meanwhile, the energy  of $\mathcal{Z}$-mode waves increases continuously,  simultaneously with $F_w(t)$. 
Note also the presence of $\mathcal O$-mode waves with $\omega^D>\omega_p$, which are generated by electromagnetic decay via $\mathcal{LZ}\longrightarrow\mathcal{O}+\mathcal{S}$,  where $\mathcal{S}$ is an ion acoustic wave, as will be shown in a forthcoming paper (see also \cite{Krafft2024}). 
The increase with time of $F_w(t)$ is mainly due to nonlinear decay that transports wave energy to smaller and more oblique wavevectors $\bf k$ until the final generation of $\mathcal Z$-mode waves with frequencies $\omega_{c\mathcal{Z}}<\omega^D\lesssim\omega_p$, where $\omega_{c\mathcal{Z}}$ is the $\mathcal Z$-mode's cutoff frequency (\textcite{Krafft2024}, \textcite{Polanco2025}). 

\subsection{Randomly inhomogeneous magnetized plasma}
\begin{figure}[h]
    \centering
    \includegraphics[width=0.5\textwidth]{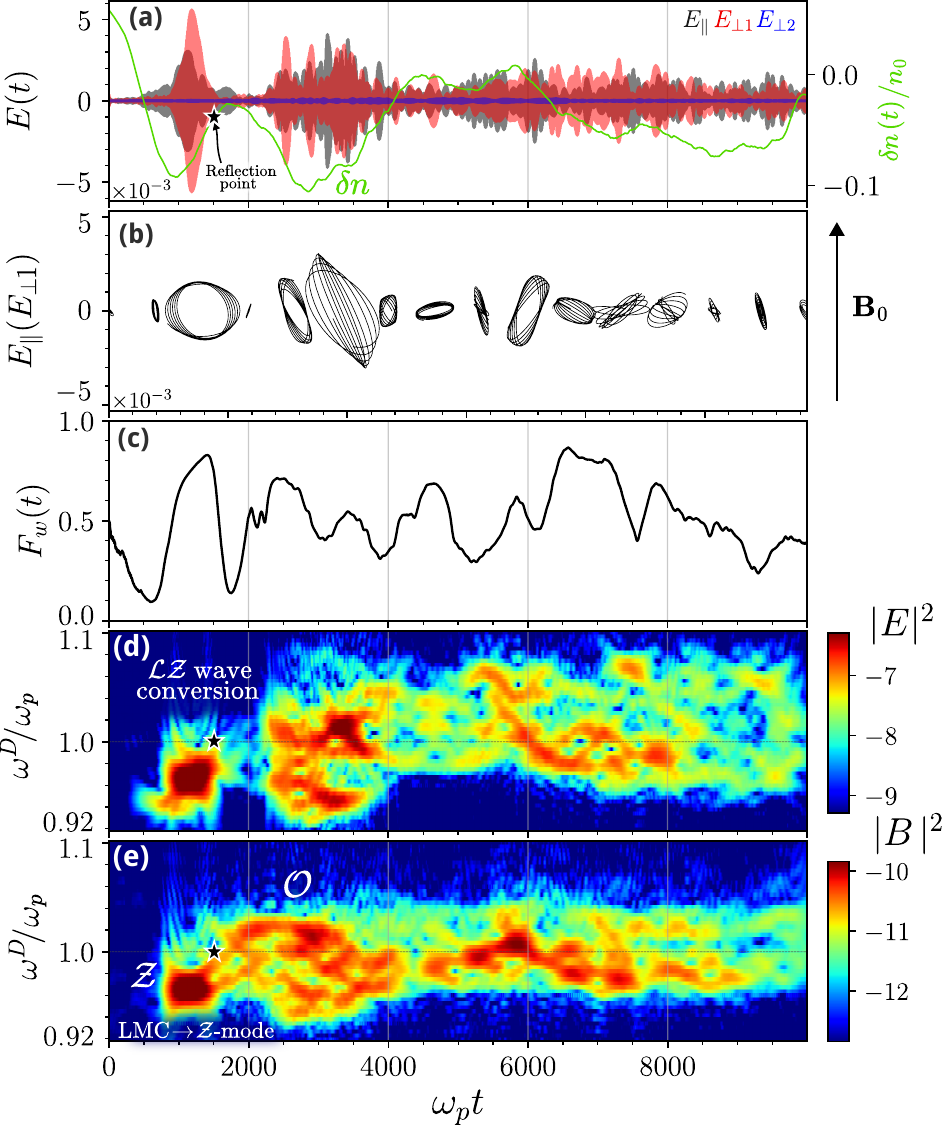}
    \caption{Typical example of waveform, together with hodograms, polarization ratios and  wave energies, in a randomly inhomogeneous and weakly magnetized plasma with $\Delta N=0.05$ and $\omega_c/\omega_p=0.07$,  for $0\leq\omega_pt\leq10,000$. 
    (a) Waveform of electric fields $E_\parallel(t)$ (gray), $E_{\perp1}(t)$ (red), and $E_{\perp2}(t)$ (blue), filtered in the frequency interval $[0.9,1.1]\omega_p$; the applied density fluctuations $\delta n/n_0$ are superimposed (green lines); a reflection point is indicated by a black star, as well as in (d) and (e). 
    (b) Hodograms $E_\parallel(E_{\perp1})$ calculated in equidistant time windows of $60\omega_p^{-1}$. 
    (c) Time variation of the polarization ratio $F_w(t)$ of the waveform. 
    (d) Electric energy spectra $|E(\omega^D,t)|^2$  in the map ($\omega_pt,\omega^D/\omega_p$). 
    (e) Magnetic energy spectra $|B(\omega^D,t)|^2$  in the map ($\omega_pt,\omega^D/\omega_p$). Both spectra are calculated using wavelet transforms.  Electromagnetic modes $\mathcal{O}$  and  $\mathcal{Z}$ are indicated by labels. All fields and  energies are presented in arbitrary units. Color bars are in logarithmic scales. }
    \label{fig2}
\end{figure}

Figure \ref{fig2} shows a typical waveform in a randomly inhomogeneous and magnetized plasma with $\Delta N=0.05$ and $\omega_c/\omega_p=0.07$. 
Contrary to the homogeneous plasma case, $E_{\parallel}(t)$ and $|E_{\perp 1}(t)|\gg |E_{\perp 2}(t)|$ reach similar intensities almost all the time (Figure \ref{fig2}a), whereas most intense wavepackets are trapped into density depletions (\cite{KrafftVolokitin2021}). 
The polarization ratio $F_w(t)$ exhibits strong enhancements up to $0.85$ (Figure \ref{fig2}c), with a mean $\langle F_w(t)\rangle_t\simeq0.49$ (averaged over  $0\lesssim\omega_p\lesssim 10,000$) around 2 times larger than in a homogeneous plasma. 
Similarly, hodograms reveal  more complex elliptical structures and rarely linear polarizations, compared with the case with  $\Delta N=0$ (Figure \ref{fig2}b). 
The spectral electric and magnetic energies show  strong electromagnetic  emissions at early times (Figures \ref{fig2}d-e), in the frequency range $0.96\lesssim\omega^D/\omega_p\lesssim0.98$, followed at $2000\lesssim\omega_pt\lesssim4000$ (corresponding to a density depletion, as evidenced in Figure \ref{fig2}a by green lines) by a wide frequency spectrum of  quasi-electrostatic  $\mathcal{LZ}$ waves of  significant energy (Figure \ref{fig2}d), which are scattered on density fluctuations and experience refraction, reflection and mode conversion (LMC), while $\mathcal Z$-mode emission at $\omega\lesssim\omega_p$ is enhanced. 

Let us focus in more detail on the trapped $\mathcal{LZ}$ wavepacket observed within $500\lesssim\omega_pt\lesssim1500$ (Figure \ref{fig2}a). 
At $500\lesssim\omega_pt\lesssim1000$, its energy is mostly carried by the parallel field $E_\parallel(t)$ generated by the beam. 
When it is reflected on the gradient of a density hump at $\omega_p t\simeq1500$ (see the black stars in Figures \ref{fig2}a,d-e indicating the reflection point), a large  perpendicular field $|E_{\perp1}(t)|>E_\parallel(t)$ is generated. 
Meanwhile, the corresponding hodogram exhibits quasi-circular polarization (Figure \ref{fig2}b). 
Moreover, a strong increase in electric and magnetic energies occurs simultaneously at $\omega^D/\omega_p\simeq0.965$, in conjunction with the reflection process (Figures \ref{fig2}d-e), which is the signature of  electromagnetic $\mathcal Z$-mode wave emission by linear conversion of $\mathcal{LZ}$ waves at constant frequency. 
Meanwhile $F_w(t)$ reaches large values up to  $0.85$, confirming a correlation between the  generation  of $\mathcal Z$-mode waves via LMC and the appearance of large polarization ratios. 

To conclude, in a weakly magnetized homogeneous plasma where $\mathcal{LZ}$ wave turbulence is generated by an electron beam,  polarization ratios grow slowly with time due to nonlinear decay of $\mathcal{LZ}$ waves. 
In randomly inhomogeneous plasmas, polarization ratios increase rapidly due to the fast linear interactions of wave turbulence with random density fluctuations, leading to scattering and conversion of $\mathcal{LZ}$ waves to $\mathcal Z$-mode waves. 
Let us now confirm these observations with statistical studies.

\section{Statistical study of polarization ratios}
\subsection{Distributions and their dependencies on beam-plasma parameters}

\begin{figure*}
    \centering
    \includegraphics[width=0.8\textwidth]{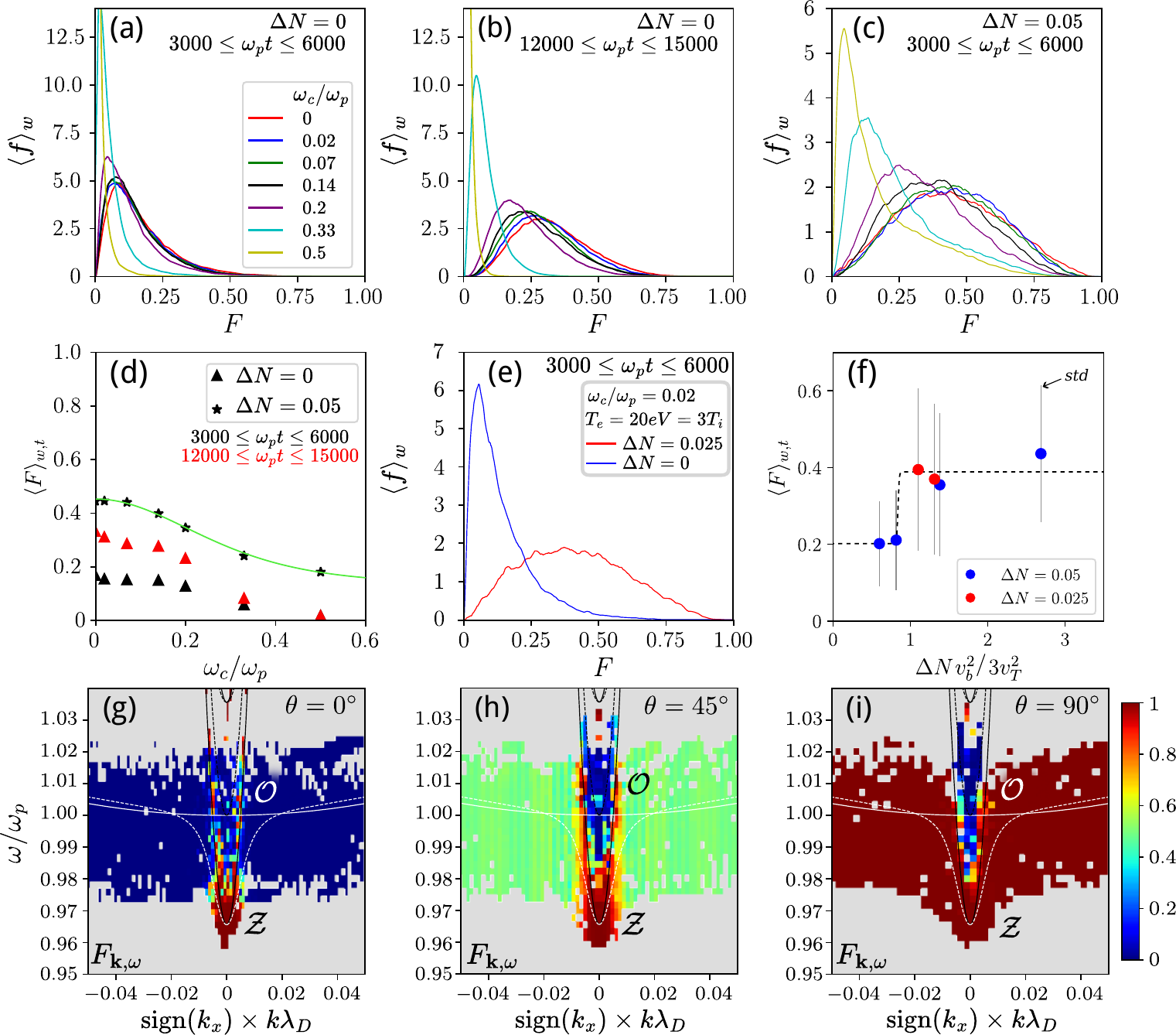}
          \caption{Distributions of polarization ratios as a function of beam and plasma parameters. 
          (a-c) Distributions $\langle f\rangle_w$ calculated using $N_w=1000$ waveforms with frequencies filtered in the range $[0.9,1.1]\omega_p$, in magnetized plasmas with $0\leq \omega_c/\omega_p\leq 0.5$, in the time intervals $3000\leq\omega_pt\leq6000$ (a and c) and $12,000\leq\omega_pt\leq15,000$ (b), for $\Delta N=0$ (a-b) and $\Delta N=0.05$ (c). 
          (d) Variations of $\langle F\rangle_{w,t}$ with $\omega_c/\omega_p$, for $\Delta N=0$ (triangles) and $\Delta N=0.05$ (stars), in the ranges $3000\leq\omega_pt\leq6000$ (black) and $12,000\leq\omega_pt\leq15,000$ (red); stars are fitted by a $\alpha+\beta\cosh^{-2}(\gamma\omega_c/\omega_p)$ curve for  $\omega_c/\omega_p\leq0.5$  (green).  (e) Distribution $\langle f\rangle_w$ calculated similarly to (a), for simulations  with $v_b=12.7v_T$, $T_e=20eV=3T_i$, $\omega_c/\omega_p=0.02$, $\Delta N=0$ (blue) and $\Delta N=0.025$ (red). 
          (f) Variation of $\langle F\rangle_{w,t}$ with $\Delta N v_b^2 /3v_T^2$, for $\Delta N=0.025$ (red points) and $\Delta N=0.05$ (blue points), with beam velocities in the set $v_b/v_T=6,7,9,11.5,12.7$; the gray vertical lines indicate the standard deviations (std) of distributions; the curve fitting the color points is represented by a black dashed line. 
          (a-f) In order to ignore noise, only times $t$ of wavepackets $\mathbf{E}(t)$ corresponding to spectral energies $|E(\omega,t)|^2$ greater than $|E(\omega,t)|_{max}^2/1000$ are retained when building distributions. 
          (g-i) Polarization ratio $F_{\mathbf k,\omega}$ (\ref{Fk}) calculated using the Fourier components $\mathbf E_{\mathbf k,\omega}$, in a randomly inhomogeneous and magnetized plasma with $\Delta N=0.025$ and $\omega_c/\omega_p=0.07$, in the time range $3500\leq\omega_p t\leq6500$. 
          $F_{\mathbf k,\omega}$ is presented in the map $(\textnormal{sign}(k_x)k\lambda_D,\omega/\omega_p)$, for 3 different propagation angles $\theta = 0^\circ,45^\circ$ and $90^\circ$; only colored regions, which correspond to  spectral electric  energies satisfying $|E_\mathbf{k}|^2>|E_\mathbf{k}|^2_{max}/1000$, are shown (other regions are represented in light gray). $\mathcal{O}$- and $\mathcal{Z}$-mode waves are indicated by labels.
          The color bar holds for (g-i). 
          A large simulation box with $L_x\times L_y=5,792\times2,896\lambda_D^2$  ($N_x=4,096$ and $N_y=2,048$ grid points) is used. 
          The theoretical dispersion curve*s of electromagnetic modes are superimposed with white and black lines for parallel (solid) and perpendicular (dashed) wave propagation. }
    \label{fig3}
\end{figure*}

In Figures \ref{fig3}a-c, the PDFs $\langle f\rangle_w$ (see section 2.3) represent the histograms of polarization ratios $F$ calculated at any time along $N_w=1000$ waveforms filtered in the frequency range $[0.9,1.1]\omega_p$ --- where, in weakly magnetized plasmas with $\Delta N\lesssim0.05$,  all $\mathcal{LZ}$ and fundamental electromagnetic wave energy is gathered --- and recorded in the intervals $3000\leq\omega_pt\leq6000$ (a and c) and $12,000\leq\omega_pt\leq15,000$ (b), in homogeneous ($\Delta N=0$) and inhomogeneous ($\Delta N=0.05$) plasmas with various $\omega_c/\omega_p$ (see caption).
The first time interval corresponds to the period when the  process of linear mode conversion (LMC) at constant frequency of $\mathcal{LZ}$ waves into electromagnetic waves is most intense (in randomly inhomogeneous plasmas with  $\Delta N\gtrsim3(v_T/v_b)^2$, as discussed below), and the second one to larger times when nonlinear three-wave decay processes are active or close to be achieved (in quasi-homogeneous plasmas with $\Delta N\lesssim3(v_T/v_b)^2$, see also hereafter). 
Figure \ref{fig3}a shows that $\langle f\rangle_w$ peaks near $F_m\simeq0.1$ (polarization ratio at the maximum of $\langle f\rangle_w$) for $\omega_c/\omega_p<0.2$ and covers the range $F\lesssim0.6$, with $\langle F\rangle_{w,t}\simeq0.17$; at larger $\omega_c/\omega_p\geq0.2$, the maxima of $\langle f\rangle_w$ correspond to much smaller $F_m$, due to the Lorentz force.
At later times  (Figure \ref{fig3}b), $F_m$ is significantly larger for all $\omega_c/\omega_p$, due to the occurrence of nonlinear three-wave interactions (see discussion below); the largest $F_m\simeq0.3$ are reached for $\omega_c/\omega_p<0.2$. 
The dependence of $\langle f\rangle_w$ on $\omega_c/\omega_p<0.2$ is weak, but much more important for $\omega_c/\omega_p\geq0.2$ (moderately magnetized plasmas).
For $\Delta N=0.05$ (with $\Delta N\gtrsim 3(v_T/v_b)^2$, see Figure \ref{fig3}c), random density fluctuations are responsible for the significant increase of $F_m$, compared with the homogeneous plasma case   (Figure \ref{fig3}a). 
For $\omega_c/\omega_p<0.2$, all PDFs $\langle f\rangle_w$ extend up to $F\simeq1$ and peak near $F_m\simeq0.4$ whereas for $\omega_c/\omega_p\geq0.2$, $F_m$ is significantly smaller.
Note that for such inhomogeneous plasmas the distributions $\langle f\rangle_w$  change little over time (not shown here). 
It appears that $F_m$ mainly increases due to $\mathcal{LZ}$ wave interactions with density fluctuations, whereas it is weakly (significantly) impacted by plasma magnetization for $\omega_c/\omega_p<0.2$ ($\omega_c/\omega_p\gtrsim0.2$). 

Figure \ref{fig3}d shows the decrease with  $\omega_c/\omega_p$ of the mean  $\langle F\rangle_{w,t}$ of all  distributions presented  in Figures \ref{fig3}a-c. 
At early times, $\langle F\rangle_{w,t}$ reaches up to 3 times larger values for $\Delta N=0.05$ than for $\Delta N=0$;  at late times, when electrostatic decay is completed, $\langle F\rangle_{w,t}$ remains  at any time significantly  smaller than in inhomogeneous plasmas, confirming that, whatever the plasma magnetization is, LMC is more efficient and fast  than electrostatic decay to generate waves with large polarization ratios. 
Note that, when the plasma is randomly inhomogeneous, $\langle F\rangle_{w,t}$ follows a dependence of the form $\cosh^{-2}(\omega_c/\omega_p) $ (the fit satisfies $R^2>0.99$), which can be approximated by a parabolic profile  for $\omega_c/\omega_p\lesssim0.2$, where the dependence of  $\langle F\rangle_{w,t}$ with magnetization is very weak. For homogeneous plasmas where nonlinear decay dominates, the tendency is linear at $\omega_c/\omega_p\leq0.2$.

Figure \ref{fig3}e shows the distributions $\langle f\rangle_w$ for the set of parameters corresponding to a significantly lower electron temperature $T_e=20eV$, with $T_e/T_i=3$, $\omega_c/\omega_p=0.02$ and $\Delta N=0, 0.025$. 
The  PDFs obtained for homogeneous and inhomogeneous plasmas correspond well to the distributions with $\omega_c/\omega_p<0.2$ in Figures \ref{fig3}a and \ref{fig3}c, respectively, showing that $T_e$ and $T_e/T_i$ weakly influence on the polarization ratio. 

Figure \ref{fig3}f shows the variation of the mean $\langle F\rangle_{w,t}$  (color points), with its corresponding standard deviations $\langle(\langle F\rangle_{w,t}-F_w(t))^2\rangle_{w,t}^{1/2}$ (gray vertical lines), as a function of $\Delta Nv_b^2/3v_T^2$, i.e. for different $v_b/v_T$ and $\Delta N$. 
Whereas $\langle F\rangle_{w,t}\simeq0.2$ for $\Delta N\lesssim3(v_T/v_b)^2$, a sharp increase of  $\langle F\rangle_{w,t}$ is observed at $\Delta N\simeq3(v_T/v_b)^2$, reaching $\langle F\rangle_{w,t}\simeq0.4$ for $\Delta N\gtrsim3(v_T/v_b)^2$ (as found in Figure \ref{fig3}c). 
This is due to a significant increase of LMC efficiency, which can be used to estimate $\Delta N$ in the solar wind (see below).

Figures \ref{fig3}g-i present the  distributions $F_{\mathbf k,\omega}$ in the $(\textnormal{sign}(k_x) k\lambda_D,\omega/\omega_p)$ map, calculated in the time range $3500\leq\omega_p t\leq6500$ using a high-resolution simulation plane with  $\delta k_x=0.001\lambda_D^{-1}$ and $\delta k_y=0.002\lambda_D^{-1}$, for three propagation angles $\theta=0^\circ,45^\circ$ and $90^\circ$ ($\cos\theta=\mathbf k\cdot\mathbf B_0/kB_0$). 
For $\theta<90^\circ$, one observes that the largest $F_{\mathbf k,\omega}\simeq 1$ are reached near the cutoff frequency $\omega_{c\mathcal Z}$ of  $\mathcal Z$-mode waves, in the spectral domain where $k\lambda_D\lesssim 0.005-0.008$; at $\theta = 90^\circ,$ $F_{\mathbf k,\omega}\simeq 1$ for $\mathcal{Z}$-mode waves with frequencies $\omega_{c\mathcal Z}\lesssim\omega\lesssim\omega_p$ and $k\lambda_D\lesssim 0.02$. 
This confirms the link between large polarization ratios and $\mathcal{Z}$-mode waves' excitation.
On the other hand, for $\mathcal O$-mode waves, which are lying in the spectral range corresponding to $k\lambda_D\lesssim 0.005$ and $\omega\gtrsim\omega_p$,  $F_{\bf k,\omega}$ varies between $0$ and $1$ depending on $\theta$, contrary to $\mathcal Z$-mode waves at $\omega_{c\mathcal Z}$, with $F_{\bf k,\omega}\simeq 1$ for any $\theta$.
Moreover, waves with $k\lambda_D\gtrsim0.02$ are quasi-electrostatic $\mathcal{LZ}$ waves with polarization $F_{\mathbf k,\omega}\simeq0$ at $\theta\simeq0^\circ$, which satisfy $E_\parallel/E_\perp\simeq k_\parallel/k_\perp$ (i.e. $F(\theta)\simeq\sin^2\theta$, as we can observe).
Note that $\mathcal X$-mode waves are only weakly emitted (not shown here), so that they are of no interest in this study.

\subsection{Discussion}
At this stage, several conclusions can be stated. 
First, random density fluctuations $\delta n$ are primarily responsible for large polarization ratios, due to their fast  interactions with $\mathcal{LZ}$ waves, which are scattered and linearly converted by LMC into electromagnetic $\mathcal Z$-mode waves excited below $\omega_p$ down to their cutoff frequency $\omega_{c\mathcal{Z}}$ (\cite{Krafft2025}), where polarization ratios can reach $F\simeq1$. 
On the other hand, in homogeneous plasmas, $\mathcal{LZ}$ wave energy can be transported toward smaller and more oblique wavevectors $\mathbf{k}$ by electrostatic decay, on much longer time scales (\cite{Polanco2025}), so that polarization ratios increase slowly with time, reaching ultimately lower values than in inhomogeneous plasmas.

Second, the polarization ratio $F_m$ (at which the PDF $\langle f\rangle_w$ is maximum) and $\langle F\rangle_{w,t}$ (averaged over time and waveforms) decrease with increasing $\omega_c/\omega_p$, for $\Delta N=0$ and $\Delta N>0$, due to the Lorentz force and to the fact that the $\mathcal Z$-mode cutoff frequency $\omega_{c\mathcal Z}\simeq \omega_p-\omega_c/2$ decreases with increasing $\omega_c/\omega_p$. 
Indeed, density fluctuations cannot scatter $\mathcal{LZ}$ waves down to frequencies smaller than $\omega \simeq(1-\Delta N)\omega_p$, and electrostatic decay leads ultimately to waves with  frequencies $\omega_{c\mathcal Z}<\omega_{min}\leq\omega\lesssim \omega_p$ above a limit $\omega_{min}$ as well (\cite{Polanco2025}). 
Consequently, for all $\Delta N$, when $\omega_c/\omega_p$ increases, $\mathcal Z$-mode waves are more difficult to generate at $\omega\simeq\omega_{c\mathcal Z}$ by beam-driven $\mathcal{LZ}$ wave turbulence, so that $F_m$ decreases. 
More precisely,  $\mathcal Z$-mode waves can only be generated  at $\omega\simeq\omega_{c\mathcal Z}$ via linear transformations of $\mathcal{LZ}$ waves on density fluctuations if $\Delta N\gtrsim\omega_c/2\omega_p$. 
 Therefore, solar wind observations of circularly polarized  $\mathcal Z$-mode waves can provide a low estimate of $\Delta N$, provided that the satellite measures $\omega_c/\omega_p$.
 Moreover, in plasmas with $\Delta N\lesssim\omega_c/2\omega_p$, polarization ratios should be lower compared to those with  $\Delta N\gtrsim\omega_c/2\omega_p$, as the radiated $\mathcal Z$-mode waves cannot be excited down to their cutoff  frequency. This can be observed in Figure \ref{fig3}d where, for $\Delta N=0.05$, $\langle F\rangle_{w,t}$ starts decreasing at $\omega_c/\omega_p\simeq2\Delta N =0.1$.

Third, for larger beam velocities $v_b$, beam-driven $\mathcal{LZ}$ waves exhibit smaller $k_b\lambda_D=v_T/v_b$, and are therefore more affected by the presence of density fluctuations $\delta n$. 
Such regime was studied by the authors (\textcite{Krafft2013}) who considered beam-driven electrostatic wave turbulence in one-dimensional plasmas where $\Delta N\lesssim3(v_T/v_b)^2$ and $\Delta N\gtrsim3(v_T/v_b)^2$, respectively (see also \cite{Ryutov1969}). 
In the former case, $\mathcal{LZ}$ waves are insensitive to  density fluctuations but can experience electrostatic decay and, in the latter one, they are transformed on $\delta n$ via various linear effects (reflection, refraction, scattering, trapping, tunneling, etc), reducing the occurrence of nonlinear decay to plasma regions where density turbulence is weak.

Moreover, space observations in type III solar wind regions showed that polarization ratios obtained from recorded waveforms increase  sharply at $v_b\simeq v_0$ ($v_0\simeq0.08c$) (\cite{Malaspina2011}, \cite{Graham2012}, \cite{GrahamCairns2013,GrahamCairns2014}), from polarization ratios $F\simeq0.1$ at $v_b\lesssim v_0$ to $F\simeq0.45$ at $v_b\gtrsim v_0$; authors suggested that electrostatic decay or scattering on density fluctuations could explain this dependence. 
In Figure \ref{fig3}f,  one observes that the boundary $v_b\simeq v_0$ is associated to the condition $\Delta N\simeq 3(v_T/v_b)^2$. 
At $v_b\lesssim v_0$ ($v_b\gtrsim v_0$), the inequality  $\Delta N\lesssim 3(v_T/v_b)^2$ ($\Delta N\gtrsim 3(v_T/v_b)^2$) is satisfied, corresponding to lower (higher) values of $F$, i.e. to the prevalence of electrostatic decay or LMC, respectively. 
This  result allows us to diagnose the average level of density fluctuations of solar wind plasmas, by calculating the polarization ratios on the basis of recorded waveforms. For example, using $\Delta N\simeq3(v_T/v_b)^2$ at $v_b\simeq v_0\simeq0.08c$ in a plasma near 1 au ($T_e\simeq10eV$) as in  \textcite{Malaspina2011}, we obtain that $\Delta N\simeq 0.01$, which corresponds well to levels of density turbulence expected at such distance. 
Note that a rough estimate of $\Delta N$ was performed in the latter work (which does not contradict our result), using however a simplified model of monochromatic Langmuir wave scattering on a density gradient. 
On the contrary, our simulations involve random density fluctuations and magnetization, and our method provides a  more precise determination of $\Delta N$. 
Furthermore, \textcite{GrahamCairns2013} classified the waveforms based on their  polarization ratios $  \langle F_w(t)\rangle_t$, designating those with $ \langle F_w(t)\rangle_t>0.2$ as corresponding to small-$k$ waves at the final stage of electrostatic decay. 
By comparing Figures \ref{fig3}a and \ref{fig3}b (where decay is predominant), we observe that polarization ratios grow from $0.1$ at early times, when decay starts to develop, to  $0.25$, when decay has transported wave energy toward small $k$-scales. This trend supports, at least in part, the classification scheme of \cite{GrahamCairns2013}.
\begin{figure*}[!hbtp]
    \centering
    \includegraphics[width=0.75\textwidth]{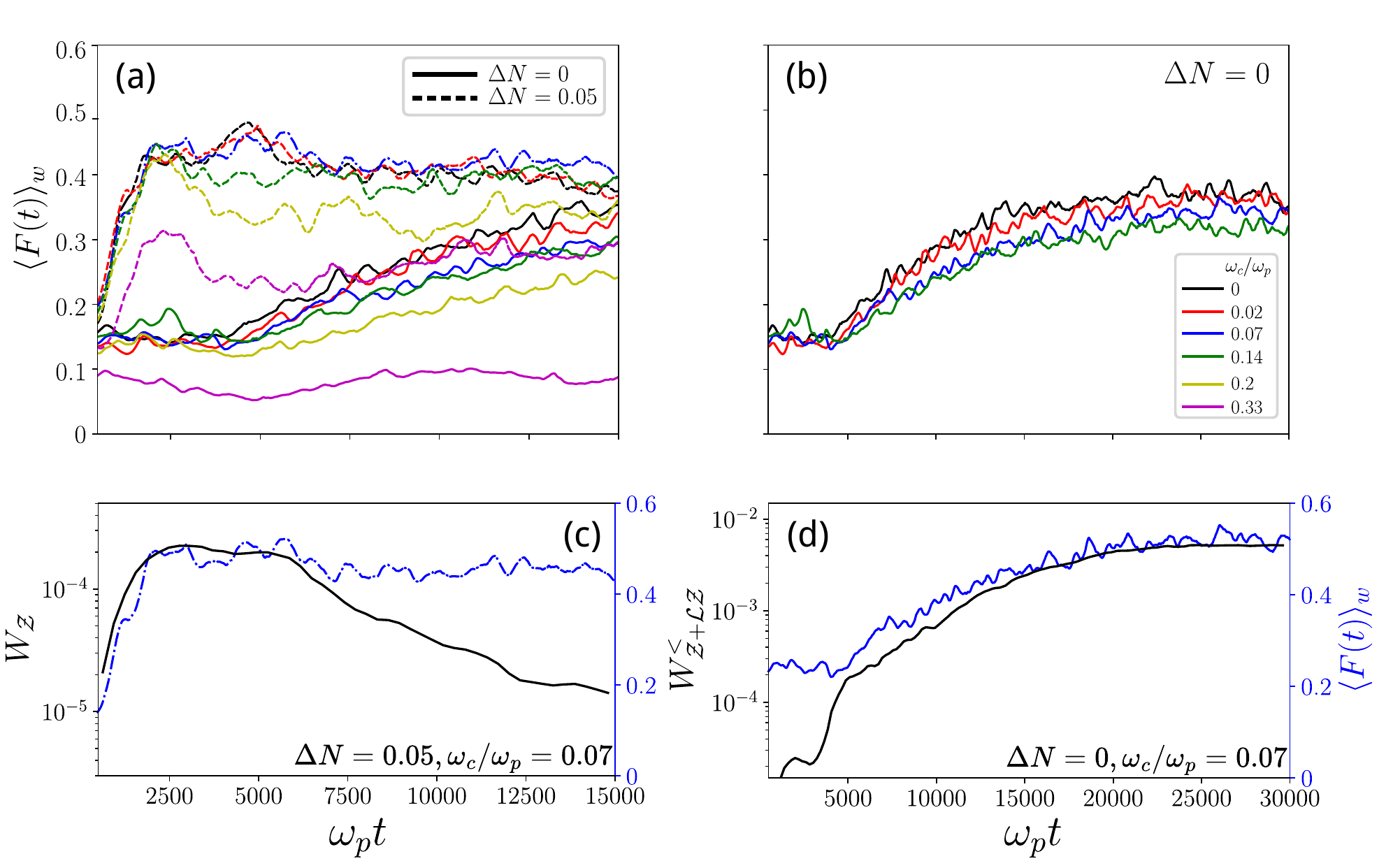}
   \caption{(a-b) Variations with time of $\langle F(t)\rangle_w$ (\ref{F2}) averaged over $N_w=1000$ waveforms  filtered in the frequency range $[0.9,1.1]\omega_p$, in plasmas with various $\Delta N$ and $\omega_c/\omega_p$. 
    (a) Homogeneous  ($\Delta N=0$, solid lines) and randomly inhomogeneous  ($\Delta N=0.05$, dashed lines) plasmas with different  $\omega_c/\omega_p \leq 0.33$ (see legend in (b)); the time range is  $\omega_pt\leq15,000$. 
    (b) Homogeneous plasmas with weak magnetization $\omega_c/\omega_p \leq 0.14$, in the very long time range $\omega_p t\leq30,000$. 
    (c) Variation with time of $\langle F(t)\rangle_w$, for $\omega_c/\omega_p$=0.07 and $\Delta N=0.05$ (solid blue line replicating the dashed blue line in (a), right axis), to which the time variation of the $\mathcal{Z}$-mode wave energy $W_{\mathcal{Z}} $ is superimposed (black dashed line, left axis).  (d) Variation with time of $\langle F(t)\rangle_w$, for $\omega_c/\omega_p$=0.07 and $\Delta N=0$ (solid blue line as in (b), right axis), to which the time variation of the energy $W_{\mathcal Z+\mathcal{LZ}}^< $  of small-$k$ $\mathcal{LZ}$ and $\mathcal{Z}$-mode waves is superimposed (black  dashed line, left axis). Energies are normalized by the initial beam energy.}
    \label{fig4}
\end{figure*}

We have shown that both LMC and decay processes lead to the increase of polarization ratios, but that LMC is clearly the fastest and most efficient  one for beam-plasma systems  with $\Delta N\gtrsim 3(v_T/v_b)^2$. 
Otherwise, when  $\Delta N\lesssim 3(v_T/v_b)^2$, distributions of polarization ratios are closer to those obtained in homogeneous plasmas where $F_m\sim0.2-0.3$ at large times. 
Therefore, the increase of polarization ratios with $v_b$ is strongly related to the condition $\Delta N\gtrsim 3(v_T/v_b)^2$ and the efficiency of LMC, rather than to electrostatic decay of $\mathcal{LZ}$ waves. 
Furthermore, the sharp growth of polarization ratios occurring at $\Delta N\simeq 3(v_T/v_b)^2$ can be used to measure the average level of random density fluctuations or the beam drift velocity in beam-plasma systems, as we did above with space observations at 1 au. 
In addition, measurements of polarization ratios' distributions as presented in Figures \ref{fig3}a-d can also provide useful estimates, as the level of density turbulence $\Delta N$ and the plasma magnetization $\omega_c/\omega_p$.

Finally, the electron temperature and the  ratio $T_e/T_i$, when varied in the ranges $20eV\leq T_e\leq 200eV$ and $3\leq T_e/T_i\leq 10$, weakly affect the polarization ratios in plasmas with parameters relevant to type III radio bursts. 
However, in a homogeneous plasma with $T_e=20eV$, the beam-plasma instability growth rate and the wave energy saturation level of $\mathcal{LZ}$  waves are smaller than in a plasma with larger temperatures $T_e$, so that the dynamics of nonlinear wave processes is significantly slower.

\subsection{Correlations between polarization ratios and $\mathcal Z$-mode waves}
Preceding conclusions can be completed and reinforced by studying the time dependence of the polarization ratio $\langle F(t)\rangle_w$  for various $\omega_c/\omega_p$, in plasmas with $\Delta N\lesssim 3(v_T/v_b)^2$ and $\Delta N\gtrsim 3(v_T/v_b)^2$. 
Figures \ref{fig4}a-b show the time variations of $\langle F(t)\rangle_w$, averaged over $N_w=1000$ waveforms recorded in the time ranges $0\leq\omega_p t\leq15,000$ (a) and $0\leq\omega_pt\leq30,000$ (b), with $\Delta N=0.05$ and $\Delta N=0$, respectively,  for $\omega_c/\omega_p\leq0.33$ (only simulations with $v_b=12.7v_T\simeq0.25c$ are used).  
In inhomogeneous plasmas with $\Delta N\gtrsim3(v_T/v_b)^2$ (Figure \ref{fig4}a), $\langle F(t)\rangle_w$ increases rapidly within the time interval $\omega_pt\lesssim2000$ for all $\omega_c/\omega_p$ and, for $\omega_c/\omega_p<0.2$, remains at a quasi-constant level around $\langle F(t)\rangle_w\simeq0.45$ during a long period ($3000\lesssim\omega_p t\lesssim15,000$); however,  the  maxima and  saturation levels reached by $\langle F(t)\rangle_w$ decrease with increasing  $\omega_c/\omega_p\geq0.2$. 
Indeed, linear interactions of $\mathcal{LZ}$ waves with density fluctuations rapidly transform and convert them to small-$k$ electromagnetic waves via LMC, increasing their polarization ratios very fast.
On the other hand, for $\Delta N\lesssim3(v_T/v_b)^2$ (Figure \ref{fig4}b), $\langle F(t)\rangle_w$ slowly grows during the time interval $\omega_pt\lesssim15,000$, for all $\omega_c/\omega_p$. 
Indeed, for $\omega_c/\omega_p<0.2$, $\langle F(t)\rangle_w$ increases from $0.15$ up to $0.3$ within $5000\leq\omega_p t \leq20,000$  (Figure \ref{fig4}b), which is at least an order of magnitude longer than the time needed to reach  maximum in a randomly inhomogeneous plasma with $\Delta N=0.05$. 
This is because,  when  $\Delta N\lesssim3(v_T/v_b)^2$, nonlinear electrostatic decay of $\mathcal{LZ}$ waves ---much slower than LMC--- is the dominant wave process.
Likewise, at $\omega_pt\simeq 30,000$, $\langle F(t)\rangle_w$ saturates at noticeably smaller values than for $\Delta N=0.05$ (for any  $\omega_c/\omega_p$),  as linear mode conversion transfers wave energy to smaller wavevectors than electrostatic decay.

In Figures \ref{fig4}c,d we replicate the time variations of $\langle F(t)\rangle_w$ presented in (a) and (b) for $\omega_c/\omega_p=0.07$,  to which we superimpose (black dashed  lines) 
the corresponding  time variations of $W_{\mathcal{Z}}$ and $W_{\mathcal Z+\mathcal{LZ}}^< $ which represent, respectively, (i) the electromagnetic energy  of $\mathcal{Z}$-mode waves (as calculated in \cite{Krafft2025})  and (ii)  the energy  carried by small-$k$  magnetized $\mathcal{LZ}$ waves  (as $\mathcal{LZ}^{(3)}$ and $\mathcal{LZ}^{(4)}$, see section \ref{section homogeneous magnetized plasma}) and by $\mathcal{Z}$-mode waves  generated at the ultimate stage of decay (as calculated in \cite{Polanco2025}). One can observe  that blue and black curves follow the same  variation within the time range where LMC is dominant ($\omega_p t\lesssim7000$ for (c)) or during the late stage of decay ($\omega_p t\gtrsim7000$ for (d)). This shows the strong correlation between the growth of polarization ratios and of  wave energies $W_{\mathcal{Z}}$ and $W_{\mathcal Z+\mathcal{LZ}}^< $. 

\begin{figure}[h]
    \centering
    \includegraphics[width=0.35\textwidth]{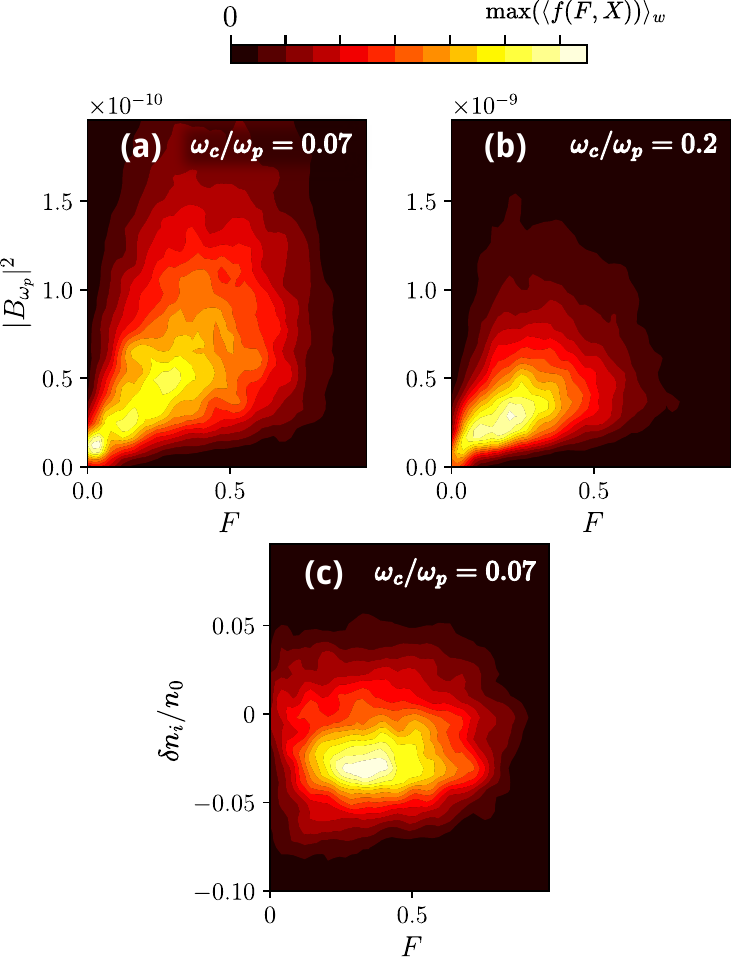}
    \caption{Distributions $\langle f(F,X)\rangle_w$ averaged over $N_w=1000$ waveforms in the time interval $1000\leq\omega_p t\leq4000$ (with frequencies filtered in the range $[0.9,1.1]\omega_p$), represented in the map $(F,X)$. 
    (a-b) $X=|B_{\omega_p}|^2$,  $\Delta N=0.05$,  $\omega_c/\omega_p=0.07$ (a) and $\omega_c/\omega_p=0.2$ (b). 
    (c) $X=\delta n_i/n_0$, $\Delta N=0.05$ and $\omega_c/\omega_p=0.07$. 
    The upper limit of the linear color bar represents, for each distribution, the maximal value  of $\langle f(F,X)\rangle_w$, which is different  for each panel. 
  In order to ignore noise, only wavepackets whose spectral energy $|E(\omega_p,t)|^2$ is greater than $|E(\omega_p,t)|_{max}^2/100$ are considered when building distributions. Energies are in arbitrary units.}
    \label{fig5}
\end{figure}

Finally, let us study the dependence of the magnetic energy carried by electromagnetic waves (mostly by $\mathcal{Z}$ and $\mathcal{O}$-mode waves) with polarization ratios, in a plasma with $\Delta N\gtrsim3(v_T/v_b)^2$. 
Figures \ref{fig5}a-b present the PDFs $\langle f(F,|B_{\omega_p}|^2)\rangle_w$, obtained using $N_w=1000$ waveforms filtered at $\omega\simeq\omega_p$, in the time range $1000\leq\omega_p t\leq4000$ (where LMC is most efficient), for $\Delta N=0.05$ and  $\omega_c/\omega_p=0.07, 0.2$. 
We observe that the magnetic energy $|B_{\omega_p}|^2$ increases quasi-linearly with $F$. 
As under used parameters most of the electromagnetic energy is carried by $\mathcal Z$-mode waves (\cite{Krafft2025}),  a clear  correlation exists between  $\mathcal Z$-mode wave emission and growth of polarization ratios. 
Note that, at later times, when $\mathcal{LZ}$ wave turbulence starts to damp (see Figure \ref{fig4}c), due mainly to the formation of a tail of accelerated beam electrons (\cite{KrafftSavoini2023}), the correlation between $|B_{\omega_p}|^2$ and  $F$, which is strong in the time interval when LMC reaches maximum efficiency, tends to disappear (not shown here). 

Figure \ref{fig5}c shows the distribution $\langle f(F,\delta n_i/n_0)\rangle_w$ as a function of $F$ and the ion density perturbation $\delta n_i/n_0=(n_i-n_0)/n_0$, in a plasma with $\Delta N=0.05$ and $\omega_c/\omega_p=0.07$.  
The distribution peaks at $-0.05\lesssim \delta n_i/n_0<0$, with most of polarization ratios satisfying $0.2\lesssim F\lesssim 0.7$, and a non negligible fraction reaching $F\gtrsim0.9$. This observation can be done for all $\omega_c/\omega_p$ studied (not shown here, see also Figure \ref{fig4}c). 
Therefore, in weakly to moderately magnetized plasmas with $\Delta N\gtrsim 3(v_T/v_b)^2$, most of  $\mathcal{LZ}$ waves of significant energy are trapped in density wells, where their frequency is shifted below $\omega_p$, enabling them to  convert at constant frequency to electromagnetic  $\mathcal{Z}$-mode waves with $\omega\lesssim\omega_p$ (see also Figures \ref{fig2}d,e). This confirms the fact that LMC is indeed the most important process of $\mathcal{LZ}$ waves' transformation into $\mathcal{Z}$-mode waves.

\section{Conclusion}

Large-scale and long-term 2D/3V PIC simulations are performed with  parameters relevant to type III radio bursts;  $\mathcal{LZ}$ wave turbulence is generated by an electron beam in a  weakly magnetized and randomly inhomogeneous solar wind plasma and radiates electromagnetic waves around the plasma frequency  $\omega_p$. 
The polarization ratios $F=|E_\perp|^2/|E|^2$ of turbulent $\mathcal{LZ}$ and electromagnetic wavepackets, which  are studied in detail,  are closely related to wave generation mechanisms and to beam and plasma properties. 
This work  provides answers to long-standing questions and offers new insights into the role of random density fluctuations in weakly magnetized plasmas with developed wave turbulence.

Statistical studies using large numbers of waveforms recorded by virtual satellites  moving in the simulation plane are performed to determine the distributions of polarization ratios $F$ and their variations with  beam and plasma parameters (beam velocity, plasma's magnetization, temperatures and average level of density fluctuations). This method, which mimics waveform recording by spacecraft in the solar wind, leads to results consistent with observations.

Plasma random density fluctuations turn out to be the key factor behind the
increase in polarization ratios, whereas high plasma magnetization $\omega_c/\omega_p\gtrsim0.2$ tends to reduce them. 
We demonstrate unambiguously that linear mode conversion at constant frequency (LMC) of $\mathcal{LZ}$ waves scattering on $\delta n$ is the most  efficient and fast process to increase polarization ratios up to $F\simeq 1$ in randomly inhomogeneous plasmas where $\Delta N\gtrsim3(v_T/v_b)^2$, due to $\mathcal Z$-mode wave emission. 
On the other hand, the nonlinear electrostatic decay ---which is a dominant process only in plasmas with weaker average levels of density fluctuations satisfying $\Delta N\lesssim3(v_T/v_b)^2$--- ultimately leads to lower values of $F$, reached after at least an order of magnitude longer time periods. 
Furthermore it is shown that, if $\Delta N\gtrsim\omega_c/2\omega_p$, large polarization ratios up to $F\simeq 1$ can be reached by electromagnetic $\mathcal{Z}$-mode waves generated by LMC near their cutoff frequency, where they exhibit left-handed  circular polarization. 

Finally, our results  provide support to estimate type III beam and plasma parameters in the solar wind. 
In particular, the average level of density fluctuations $\Delta N$ can be inferred from the dependence of polarization ratios derived from in situ  observed waveforms on beam velocities measured by satellites, by using the relation $\Delta N\simeq3(v_T/v_b)^2$ at the beam velocity where the sharp growth of polarization ratios is observed. 
Likewise, observations in the solar wind of circularly polarized $\mathcal Z$-mode waves can supply a low estimate for $\Delta N$, provided that the magnetization ratio  $\omega_c/\omega_p$ can be measured by the spacecraft, which is usually the case. 
Moreover, the dependence  of the averaged polarization ratio on $\omega_c/\omega_p$, evidenced in this work, can  provide guidance to theoretical studies.

\section{acknowledgements}
This work was granted access to the HPC computing and storage resources under the allocation 2023-A0130510106 and 2024-A017051010 made by GENCI. This research was also financed in part by the French National Research Agency (ANR) under the project ANR-23-CE30-0049-01.  C.K. thanks the International Space Science Institute (ISSI) in Bern through ISSI International Team project No. 557, Beam-Plasma Interaction in the Solar Wind and the Generation of Type III Radio Bursts. C. K. thanks the Institut Universitaire de France (IUF).

For open access purposes, a CC-BY license has been applied by the authors to this document and will be applied to any subsequent version up to the author's manuscript accepted for publication resulting from this submission.
\bigskip  
\bibliography{./articleBiblio.bib}

\end{document}